\documentstyle[12pt]{article}

\def\halft{{\textstyle {{{1}\over{2\pi\alpha'}}} }}

\def\half{{\textstyle{1\over2}}}
\def\quarter{{\textstyle{1\over4}}}
\begin{document}
\pagestyle{empty}
\begin{titlepage}

\bigskip
\hskip 3.7in\vbox{\baselineskip12pt
\hbox{PSU-TH-209}\hbox{hep-th/9906238}}
\bigskip\bigskip\bigskip\bigskip

\centerline{\large \bf Ultraviolet Limit of Open String Theory}
\bigskip\bigskip
\bigskip\bigskip

\centerline{\bf Shyamoli Chaudhuri\footnote{Current Address: 1312
Oak Dr, Blacksburg, VA 24060. shyamolic@yahoo.com}}
\medskip
\centerline{Physics Department}
\centerline{Penn State University}
\centerline{University Park, PA 16802}

\bigskip\bigskip

\begin{abstract}
\noindent We confirm the intuition that a string theory which is
perturbatively infrared finite is automatically perturbatively
ultraviolet finite. Our derivation based on the asymptotics of the
Selberg trace formula for the Greens function on a Riemann surface
holds for both open and closed string amplitudes and is
independent of modular invariance and supersymmetry. The mass
scale for the open strings stretched between Dbranes suggests a
natural world-sheet ultraviolet regulator in the string path
integral, preserving both T-duality and open-closed string
world-sheet duality. Note added (Jan 2005): Comments and related
references added.
\end{abstract}

\end{titlepage}

\pagestyle{plain} In developing a formalism for nonperturbative
string theory, it is important to clarify which properties of the
theory follow from world-sheet principles alone. Finiteness in the
ultraviolet limit is a fundamental property of the perturbative
amplitudes of an infrared finite string theory, well known to
string experts \cite{polchinskivol1}.  Unfortunately, since most
discussions of finiteness are carried out in the context of
superstring theory it is often incorrectly attributed to
supersymmetry. It has also been misconstrued as a consequence of
modular invariance--- a property of only the closed string sector
of the theory. To clarify both of these points unambiguously, we
will examine the worldsheet ultraviolet limit of the amplitudes of
the oriented bosonic string with multiple boundaries on Dbranes.
Parallel and static Dpbranes separated by a distance $r$ in a
direction orthogonal to the branevolume lead to BPS configurations
in string theory: left-moving string modes are identified with
right, upon spacetime reflection at either Dbrane. In the
superstring this breaks half of the spacetime supersymmetries.
Relative motion of the branes in a direction orthogonal to the
branevolume or, equivalently, a relative spatial rotation of the
branes in a plane embedded in the transverse space orthogonal to
the branevolume, breaks further the spacetime supersymmetries.
But, as we will see, the absence of infinities in the ultraviolet
limit for an infrared finite string theory are quite independent
of such supersymmetry breaking boundary conditions.

\vskip 0.1in Infrared finite string theories require the absence
of tachyon and tadpole states in the string spectrum
\cite{polchinskivol1}. These conditions are easily achieved by,
respectively, introducing fermionic degrees of freedom on the
world-sheet, and including nonorientable world-sheets in the path
integral. The effect of the negative mass tachyon and the zero
momentum massless states, or tadpoles, that characterize the
infrared limit of the oriented bosonic string can be exposed
explicitly in the path integral, distinguishable from other
features of the amplitude generic to any string theory. Our
derivation of the worldsheet ultraviolet asymptotics of the
bosonic string path integral, based on the Selberg trace formula
\cite{selberg}, can be applied to any string amplitude, open or
closed, and rests upon open-closed string world-sheet duality
alone. The result can in fact be inferred from previous work on
bosonic closed string amplitudes
\cite{poltorus,dhokerphong1,wolpert3,nelson}. We have simply
clarified their extension to both open and closed string
amplitudes.

\vskip 0.1in Unlike amplitudes in the closed string sector, where
an infrared divergence can always be mapped to ultraviolet and
vice versa by a symmetry of the amplitude--- modular invariance,
it is meaningful to distinguish the infrared and ultraviolet
limits of the open string sector. In an open and closed string
theory, open-closed worldsheet duality instead maps the infrared
behavior of the open string spectrum to the ultraviolet behavior
of the closed string spectrum \cite{polchinskivol1,zeta}. Let us
review this point with a discussion of the annulus diagram in
bosonic string theory. Parameterize the annulus with world-sheet
coordinates $0\le \sigma^1 \le 1$, $0\le \sigma^2 \le 1$, and
boundaries of length $l$. The tree exchange of oriented closed
strings between a pair of parallel Dpbranes with separation $r$ in
a direction, $X^{24}$, and relative velocity $v$ in a direction,
$X^{25}$, can be written in the form \cite{scpath,velocity}:
\begin{equation}
{\cal A}(r,v) = V_{26} \int_{0}^{\infty} {{dl}\over{l}}
(8\pi^2\alpha'l)^{-p/2} e^{-r^2l/2\pi\alpha'}
   [l^{-1/2} \eta(i/l)]^{-22}[{{\eta(i/l)}\over{ - \Theta_{11}(-iu/\pi,i/l)}}],
\label{eq:ampdtwo}
\end{equation}
where $v$$=$${\rm tanh} u$. Small $r$ corresponds to large $l$ or,
in terms of the inverse length parameter, ${\tilde l}$$=$$1/l$,
small ${\tilde l}$. The closed string tachyon and
\lq\lq would-be\rq\rq  dilaton tadpoles present in the
$r$$\to$$\infty$ infrared limit are exposed
by expanding the integrand in powers of $e^{-2\pi{\tilde l}}$:
\begin{eqnarray}
\lim_{{\tilde l}\to \infty} &&
{\tilde l}^{(p-24)/2} e^{-r^2/2\pi\alpha'{\tilde l}}
     {{ e^{2\pi{\tilde l}} }\over{2i{\rm Sinh}(u)}}
          \nonumber\\
 &&\times [1+e^{-2\pi{\tilde l}}(22+2{\rm Cosh}(2u)) +
    O(e^{-4\pi{\tilde l}})]  .
\label{eq:lightclosed}
\end{eqnarray}
The leading term is absent in a tachyon-free infrared finite
string theory. Note that the tension of the stretched string,
$M_W$$\sim$$r^2l/2\pi\alpha'$, gives a natural cutoff on the
$l$$\to$$0$ limit of the string path integral.

Conversely, we can expand in powers of $e^{-2\pi l}$, which exposes
the $r$$\to$$0$ ultraviolet limit of the amplitude:
\begin{eqnarray}
\lim_{l\to\infty} && l^{-(p+2)/2} e^{-r^2l/2\pi\alpha'}
  {{ e^{-u^2l/\pi} }\over{ 2i {\rm Sin}(ul) }} e^{2\pi l} \nonumber\\
  &&\times [1+ e^{-2\pi l}(22 + 2{\rm Cos}(2ul)) + O(e^{-4\pi l})].
\label{lightopen}
\end{eqnarray}
In the absence of the leading term in the expansion, in an
infrared finite string theory, the ultraviolet limit will be
automatically finite quite independent of spacetime supersymmetry.
The ultraviolet limit is then an open string abelian gauge theory
with the Goldstone boson that arises from the spontaneous breaking
of spacetime translational invariance by Dbrane boundaries
appearing at the first massive level in the open string spectrum
\cite{bachas,dkps}. The mass of the stretched string plays the
role of the classical mass of the Goldstone boson. We will explore
the analogous limit of the tree exchange amplitude of multiple
bosonic Dbranes \cite{scpath}, using the Selberg trace formalism
\cite{selberg}.

\vskip 0.1in
The theta functions in Eq.\ (\ref{eq:ampdtwo}) are the result of summing over
eigenmodes of the Laplacian on the annulus with $p$ Neumann and $24-p$
Dirichlet coordinates, and the boundary conditions: $X^{25}=0$,
$\partial_2 X^0=0$ at $\sigma^2=0$, and $X^{25}=v X^0$ at $\sigma^2=1$,
on coordinates $X^{25},X^0$. An intuitive picture of the large/small $l$
asymptotics is provided by the diffusion time of a given eigenmode on the
world sheet \cite{alvarez,mckean}. The Greens function
of the diffusion operator, $\partial_t -\Delta$, is a trace over
the eigenvalue spectrum,
$\lambda_{(m_1,m_2)}$$=$$4\pi^2(m_1^2+m_2^2{\tilde l}^2)+E_{\rm vac.}(u)$,
and $-\infty$$\le$$m_1$$\le$$\infty$, $m_2>0$:
\begin{equation}
{\rm ln}~{\rm det}' \Delta=
    - \int_0^{\infty} dt t^{-1} \int d^2 z {\sqrt{g}} G(z,z';t).
\label{diffn}
\end{equation}
Symmetrizing under $m_2\to -m_2$, a Poisson resummation expresses the
trace in the equivalent form:
\begin{eqnarray}
\sum_{m_1,m_2} &&e^{-4\pi^2(m_1^2 + m_2^2 {\tilde l}^2)t -2m_2 u }
\nonumber\\
   &&= \sum_{n_1,n_2} {{A}\over{(4\pi t)^{1/2}}}
       e^{-(n_1^2 +(n_2 -iu/\pi)^2 l^2)/4t},
\label{poiss}
\end{eqnarray}
where $A\equiv l$ is the area of the surface. For small diffusion
times only the lowest lying eigenmode is excited; this is the
small $l$ limit which, due to the stretched string background, is
the same as the large $r$ limit. For large diffusion times,
several modes can be excited together. The dominant contribution
to the path integral comes from the most excited states: heavy
closed strings, or, by world sheet duality, light open strings
\cite{dkps}.

With this picture in mind, we now consider the extension to open string amplitudes
with $b$$\ge$$3$ boundaries on Dbranes. If we wish to identify the small $l$
limit of this amplitude as the tree gravitational interaction
of $b$ Dbranes with separations of order $r$$>>$$\sqrt{\alpha}'$, the
boundaries live on distinct branes. Now the expression on the right hand side
of Eq.\ (\ref{poiss}) has a simple geometric interpretation. An arbitrary
surface
with cylindrical topology maps to a rectangular cell in the complex plane with
edges of metric length $(1,l)$. Thus the eigenvalue trace is a sum over a
length spectrum of geodesic paths: paths joining opposite edges of the unit
cell, or, by the translational symmetry, ${\rm Re } z$$\to$${\rm Re } z+1$,
the opposite
edges of two cells related by $n_1$ translations. An extension of this idea
underlies the Selberg trace formula for the Greens function of the diffusion
operator on a Riemann surface, $S$, with $b$$\ge$$3$ boundaries
\cite{selberg,mckean}.

\vskip 0.1in
Since the sums over eigenfunctions are potentially divergent we introduce
regulators. Consider the expression:
\begin{equation}
{\rm ln}{\rm det '} \Delta = - \lim_{s \to 0} {{d}\over{ds}}
\int_{\epsilon}^{\infty} {{dt t^{s-1}}\over{\Gamma(s)}}
   \int_{S} d^2 z {\sqrt {g}} [{\rm tr'} e^{-t(\Delta + m^2)}] - e^{-m^2 t},
\label{determ}
\end{equation}
where $\epsilon$ and $m^2$ are, respectively, world--sheet
ultraviolet, and infrared, regulators, the latter being necessary
only in cases where the Laplacian develops a zero mode. The
infrared limit is then regulated by Pauli Villars subtraction as
in \cite{poltorus,dhokerphong1,zeta}. We will denote the factor in
square brackets as $ {\hat G}(z,z';t)$. Notice that the
introduction of a cutoff on diffusion time gives an ultraviolet
regulator formally violates worldsheet Weyl invariance,
introducing an explicit mass scale, $M_W$, into the path integral
formalism \cite{zeta}. $M_W$ behaves like an ultraviolet cutoff
from the perspective of the worldvolume {\em effective} gauge
theory. However, this prescription preserves both T--duality,
which relates open string backgrounds with distinct numbers of
Dirichlet/Neumann coordinates, and open--closed string world sheet
duality \cite{polchinskivol1}. It should be distinguished from the
finite world sheet cutoff for general nonperturbative string
backgrounds proposed in \cite{periwal}, which would break world
sheet duality explicitly.

The Greens function, ${\hat G}(z,z';t)$, is assembled from the
basis of eigenfunctions of the Laplacian on the Riemann
surface $S_b$ with Euler character $\chi=2-b$. The divergent contribution
from the Greens function at the source, ${\hat G}(z,z;t)$, will be
understood to have been subtracted from this trace; it can be evaluated by
the same method. By the uniformization theorem \cite{bers,abikov}, we can map
such a surface to a polygon $H/\Gamma_b$, a $-9\chi$-sided {\em subdomain} of the
upper half complex plane, $H$, with constant negative curvature,
$R_g=-1$, area element, $dA$$=$$dxdy/y^2$, $y>0$, and, from the Gauss--Bonnet
theorem, constant area, $A=-4\pi\chi$. $\Gamma_b$ is the fundamental, or
covering, group of the Riemann surface. The Laplacian takes the form
$\Delta = y^{-2} (\partial_x^2 + \partial_y^2)$. By the Riemann--Roch theorem,
the shape of the surface is characterized by $-3\chi$ real parameters. We
choose these to be the Fricke-Klein moduli \cite{abikov,frickeklein,keen}:
the lengths of $b$ holes, and $2(b-3)$ interior paths on $S_b$,
closed and nonintersecting,
and geodesic with respect to the Poincare line element,
$ds^2$$=$$d\rho^2+{\rm Sinh}^2 \rho d\theta^2$.
These will be described in an example below. See, also, the discussion of
uniformization and moduli in the appendix. Thus, the length of a
geodesic path on $H$ through the points $z$, $z'$ is given by the
minimum:
\begin{equation}
\rho_H(z,z') =  {\rm min} \int_{C} {{dy}\over{y}} = {\rm min} {\rm Cosh}^{-1}
   [ 1+ {{|z-z'|^2}\over{2y y'}}]^{1/2} ,
\label{lineel}
\end{equation}
where $C$ is an arbitrary path on $H$ linking $z$, $z'$.

Thus, ${\hat G}$ on $S_b$ is composed from the basis of eigenfunctions
of the Laplacian on the Poincare upper half plane, $H$,
with the periodicity property, $\Psi(x+{\tilde x},y;q)$$=$$e^{2\pi i q
{\tilde x}}
\Psi(x,y;q)$, under shifts, ${\rm Re} z $$\to$${\rm Re } z$$+$${\tilde x}$,
where $q$ is a real parameter. For eigenvalues parameterized
as $\lambda=-(\quarter +v^2)$, with $v$ real, these take the form
\cite{eigtext}:
\begin{equation}
\Psi_{(q,\lambda)} = N_{(q,v)} y^{1/2} K_{iv}(2\pi |q|y) e^{2\pi i q x},
\label{eigs}
\end{equation}
where $K_{iv}(w)$ is a modified Bessel function. Neumann/Dirichlet boundary
conditions may be imposed by
evaluating the Greens function on the double of $S_b$, a closed orientable
Riemann surface with $b$$-$$1$ handles, and restricting to the subspace
of eigenfunctions that are, respectively, even/odd with respect to
reflection in the boundary \cite{eigtext}.

The normalization, $N_{(q,v)}$, is determined by the integration measure:
\begin{equation}
{\hat G}(z,z';t) = N_{(q,v)}^2 \int_{-\infty}^{\infty} dq \int_{-\infty}^{\infty}
        [ v {\rm Sinh} (\pi v) dv] {\sqrt {y}}{\sqrt {y'}} K_{iv}(2\pi |q| y)
   K_{iv}(2\pi |q| y'),
\label{meas}
\end{equation}
invariant under linear fractional transformations
$z'$$\equiv$$\gamma[z]$$=$${{az+b}\over{cz+d}}$
with $a,b,c,d$ real, and $ad-bc=1$, that take the Poincare upper half plane
into itself. Thus, $\gamma$ is an element of the group $PSL(2;R)$,
and the factor
in square brackets can be recognized as the group invariant measure.
Integration over the \lq\lq angular \rq\rq variable $2\pi q$ gives the
associated Legendre function, $P_{iv-\half}[{\rm Cosh}(\rho_H(z,z'))]$.
Following some simple manipulations \cite{eigtext}, the Greens function
${\hat G}$ on $H$ can be put in the form:
\begin{equation}
{\hat G}(z, z';t) = {{1}\over{(4\pi t)^{3/2}}} \int_{\rho_H}^\infty
   {{{\sqrt{2}} b e^{-b^2/4t - M^2 t} }\over{ {\sqrt{ {\rm Cosh} b -
     {\rm Cosh} \rho_H }}}} db,
\label{grhalf}
\end{equation}
where $\rho_H$ is the length of the geodesic path joining points $z,z'$, and the
integral sums over lengths, $b$, of arbitrary paths on $H$ joining $z,z'$.
The parameter $M^2=m^2 + \quarter$. The Greens function for coincident
points $z'=z$ is obtained by simply setting the argument of the Greens function
to the ultraviolet cutoff on the world-sheet, a parameter of $O(\epsilon)$.
Insight into
the physical meaning of $\epsilon$ will be obtained by considering the
full amplitude as shown below.

The precise spectrum of eigenvalues of $\Delta$ on $H$ is not
known by analytic means \cite{eigtext}. However the growth of the
density of eigenvalues for large $\lambda$ can be estimated. We
will use this estimate below to bound the behavior of the Greens
function on $S_b$ for large $\lambda$. The opposite limit of small
eigenvalues, and short geodesics, determines the $l$$\to$$0$ limit
of the amplitude \cite{wolpert3,nelson}. The Selberg trace formula
expresses the Greens function on the fundamental polygon
$H/\Gamma_b$ to that on the covering space $H$
\cite{selberg,mckean}. Some key steps are reviewed in the
appendix. The fundamental group $\Gamma_b$ of $S_b$ is the
$-3\chi$ parameter Fuchsian subgroup of linear fractional
transformations: every element is related by a similarity
transformation to a magnification, $z'$$=$$\alpha[z]$, with
generating elements constrained by the relation
$\prod_{i=1}^{3(b-2)} \gamma_i = {\bf 1}$, and ${\rm tr} \gamma >
2$ for all $\gamma$$\in$$\Gamma_b$. More details can be found in
the appendix. The magnification, $z'=\alpha[z]$, defines a
geodesic path on $H/\Gamma_b$, with length, $\rho_H(z,z')$,
defined as above, and the $-3\chi$ real parameters of the
generating group elements can be explicitly related to the
Fricke-Klein moduli of $S$ as shown in
\cite{frickeklein,keen,wolpert1}.

To gain some intuition into the moduli, consider orientable
surfaces with four holes, $C[i]$, of common
length of $O(l)$,
and two interior geodesics, homotopic to $C[1]C[2]$, $C[2]C[3]$,
of length $l_S$, $l_T$. The reader may find it helpful to
keep in mind the bosonic contribution to the tree scattering
amplitude of four D0branes in type IIA string theory. The
simpler $b$$=$$3$ pants amplitude does not exemplify the general
case since it lacks interior moduli. Let the distances
between pairs of Dbranes satisfy the inequalities, $r_{12}
\sim r_{34} \ge {\sqrt{\alpha '}}$, with $r_{13}\sim r_{14}
\sim r >>> r_{12}$. Then the $l$$\to$$0$ limit of the amplitude
is dominated by world-sheets with four long tubes of length
${\tilde l}$ attached to the respective Dbranes. The remaining
two moduli, $l_{S}$, $l_T$, can be interpreted as the
\lq\lq proper time'' for propagation of an intermediate closed
string state in, respectively, the S and T channels of the
amputated amplitude. An explicit representation of the
generating elements of the group $\Gamma_4$ in terms of PSL(2,R)
matrices, and the parameterization in terms of the six
Fricke-Klein moduli of $S_4$, is given in \cite{keen}.

Alternatively, one can use the Fenchel-Nielsen length-twist
parameterization of the interior geodesic moduli
\cite{fenchelnielsen,keen,wolpert2}. Returning to the example
above, a pair
of coordinates, $l_S$, $\theta_S$, where $\theta_S$ is the
Dehn twist of $C[1]C[2]$ relative to $C[3]C[4]$ before
identification. In the limit $l$$\to$$0$, $l_S$$\to$$0$,
the static amplitude factorizes into the product of four
disc tadpole amplitudes with emanating closed string propagators,
and a closed string propagator linking two cubic closed
string interaction vertices, spatially separated by a
distance of $O(r)$. Upon setting the branes in motion, we
can extract from this amplitude an $O(g_c^2)$ correction to
the bosonic sector of an eight-point function of
supergravitons in M theory \cite{mlcone}. Here, $g_c$ is
the closed string coupling.

Note that the presence of classical mass terms for the Goldstone
bosons in the string path integral--- in this example,
$M_W$$\sim$$O(r/\alpha')$, with $r_{13}\sim r_{14} >>> r_{12} \sim
r_{34}$, implies a lower cutoff, $l_{\rm min.}$, on the integrals
over geodesic loop lengths. Let $l_{\rm min.}$ be the shortest,
non-contractible, geodesic loop length summed in the string path
integral, a parameter of $O(\alpha'/r^2)$. Note that cutting off
the integrals over geodesic loop length at $l_{\rm min}$ $>$ $0$
amounts to a formal violation of Weyl invariance, and can be
interpreted as the introduction of a world-sheet ultraviolet
regulator, $\epsilon$, of $O(\alpha'/r^2)$. This prescription can
be compared with the observation in \cite{maldacena}, where the
role of the \lq\lq W\rq\rq  boson mass as a spacetime infrared
regulator in the bulk supergravity theory was noted. This is a
consequence of the stretched string backgrounds, since small $l$
is the same as large $r$ in the presence of the stretched strings.
With the Fenchel--Nielsen coordinates as parameterization of
conformally inequivalent classes of world sheet metrics, ${\cal
T}$$\equiv$${\cal M}_g/{\rm Conf}\times{\rm Diff}_0$, the measure
in the path integral takes the simple form
\cite{dhokerphong2,scpath}:
\begin{eqnarray}
 {\cal A} =&& V_{26} A^{(-3\chi+p)/2} (4\pi^2\alpha')^{-p/2}
    \int \prod_{m=1}^{2b-3} \prod_{n=1}^{b-3} dl_m d\theta_n \nonumber\\
  &&\times  e^{-\sum_{i \neq j} \halft r_{ij}^2 l_{j}}
   ({\rm Det}'\Delta_1)^{1/2} ({\rm Det}'\Delta_0^{(26)})^{-1/2} ,
\label{amp}
\end{eqnarray}
where the normalization of the amplitude is understood to be absorbed in
the regulated spacetime volume, $V_{26}$, and the functional determinants
are to be computed with the given boundary conditions by an extension of
the method below \cite{dhokerphong1,dhokerphong2}. An arbitrary metric
in ${\cal M}_g$ can be uniquely transformed by a conformal transformation
plus diffeomorphism continuously connected to the identity, to the fiducial
constant curvature metric $g$ \cite{polyakov,wep,dhokerphong1}.
The parameterization of ${\cal T}$ is not unique. For a different
parameterization, the measure in the path integral
is given by the determinant computed in \cite{poltorus}:
\begin{equation}
\prod_{i=1}^{-3\chi} d\tau_i [{\rm det }
  ((\chi_k)_{ef} (\chi_l)^{ef})]^{1/2}
\label{moduli}
\end{equation}
where $\chi_k^{ab} \equiv \partial_k g^{ab} - {{1}\over{2}}
g^{ab}g^{cd}\partial_k g_{cd}$, and $g_{ab}$ is the fiducial metric on some
gauge slice in the space of metrics, ${\cal M}_g$. The Fenchel-Nielsen
coordinates for ${\cal T}$ correspond to the choice for which this determinant
is trivial. The moduli dependence is then entirely contained in the lengths,
$l_i$, of $2b-3$ closed geodesic paths on $S$ \cite{keen,wolpert1},
$b-3$ of which are interior paths, and $b-3$ corresponding twist angles, $\theta_j$
\cite{fenchelnielsen,wolpert2,dhokerphong2}. Explicit examples and more
details can be found in \cite{keen}.

The Greens function on the polygon $H/\Gamma_b$ can be written as:
\begin{equation}
{\rm tr}' e^{-(\Delta+m^2)t} =
\sum_{\gamma \neq {\bf 1}} \int_{H/\Gamma_b} {{dxdy}\over{y^2}}
 {\hat G}(z,\gamma[z];t) + A(H/\Gamma)\cdot {\hat G}(z,z;t).
\label{grdomain}
\end{equation}
The identity element in the fundamental group, $\gamma[z]=z$, gives
the Greens function at the source, ${\hat G}(z,z;t)$. It has
been separated from the sum over group elements in Eq. (\ref{grdomain}).
This defines the Poincare theta series, $\Theta(t)$, described briefly in the
appendix \cite{mckean}.
Consider the {\em primitive} elements of $\Gamma_b$, $\{\delta\}$,
transformations which cannot be expressed as a power of any other
transformation. Then an arbitrary $\gamma$$\in$$\Gamma_b$ can always be
related by a similarity
transformation to an integer power of a primitive element,
$\gamma$$=$$\beta \delta^p \beta^{-1}$, for suitable integer $p$, and
element $\beta$ noncommuting with $\delta$. For surfaces with $b$ holes,
such a similarity transformation can be used to bring any primitive
element $\delta$ to the form of a magnification \cite{selberg,mckean}. It
follows that the length parameter of the element,
$\delta^p$, satisfies the relation $l(\delta^p)=pl(\delta)$. The use of the
similarity transformations then allow us to derive \cite{selberg,mckean}:
\begin{eqnarray}
{\rm ln}{\rm det'} \Delta \equiv&& - \lim_{s\to 0} {{d}\over{ds}}
             \left [ {{1}\over {\Gamma(s)}} \int_{\epsilon}^{\infty}
                     dt t^{s-1} [ \Theta(t) + A \cdot {\hat G}({\rm Cosh}
                         (l_{\rm min.});t) ] \right ]  \nonumber\\
=&&- \lim_{s\to 0} {{d}\over{ds}} \left [
      \sum_{\{\delta\}} \sum_{p=1}^{\infty}
     {{l(\delta)/2}\over{{\rm Sinh} (pl(\delta)/2)}}
   \int_{\epsilon}^{\infty} {{dt t^{s-3/2}}\over{\Gamma(s){\sqrt{\pi}} }}
            e^{-p^2 l(\delta)^2/4t -M^2t} \right ] .
\nonumber\\
\label{eq:selb}
\end{eqnarray}
Some key steps are given in the appendix. The factor in square brackets
is a generalized zeta function
$\zeta_R(s)$ \cite{selberg,mckean,hawking}. Note that the integral over
diffusion times, which we will denote by $I(pl(\delta);\epsilon, m^2)$, is
regulator dependent.

The sum over primitive elements in Eq.\ (\ref{eq:selb}) needs further
comment. The spectrum of geodesic lengths on $S_b$ is discrete, a direct
consequence of the discreteness of the eigenvalue spectrum of
the Laplacian \cite{eigtext}. The $l$$\to$$0$ limit determines the
infrared $r$$\to$$\infty$ behavior of the amplitudes and has been extensively
studied in the literature \cite{wolpert3,nelson}. Consider the double of
the Riemann surface, $S_b$, a closed oriented surface with $b-1$ handles
and $2(b-3)$$+$$b$ pairs of Fenchel-Nielsen moduli: $l_j$, $\theta_j$,
$j$$=$$1$, $\cdots$, $3(b-2)$. In the limit of a pinched cycle,
$l_j$$\to$$0$ for some $j$, the world-sheet develops a long tubular neck,
or collar, with hyperbolic metric and eigenvalue spectrum, determined
by the behavior of the minimum length core geodesic on the pinched cycle
\cite{wolpert3}. The length and twist moduli of the pinched cycle become
redundant. The eigenvalue spectrum on the {\em complement} of the surface,
with collar region excised, can then be bounded, and the divergence in
the spectrum of the collar component isolated. The result is similar to
the long cylinder limit described earlier: in either case, it comes from
the clustering of low frequency, rotationally invariant, eigenmodes at
the limit point: $\lambda$$=$$-\quarter$ \cite{wolpert3}. This analysis
of the $l$$\to$$0$ asymptotics of the trace formula can be modified in
the presence of a finite cutoff on the world-sheet. Some preliminary
steps are given in the appendix.

We will consider here the opposite limit of the amplitude, the
worldsheet infrared $l$$\to$$\infty$ regime, dominated by the
behavior of the large eigenvalues, or long geodesics. We will find
that our suggested introduction of an ultraviolet cutoff
$\epsilon$ on the $l$ integral will result in an
$\epsilon$$\sim$$O(\alpha'/r^2)$ dependent correction to the
leading behavior of $\zeta_R(s)$ in this limit. We find, however,
that this correction is exponentially damped as
$\epsilon$$\to$$0$, confirming the intuition that a perturbatively
infrared finite string theory is automatically perturbatively
ultraviolet finite when worldsheet Weyl invariance has not been
violated.

Cut off the sum over primitives for the first $N(L)$ primitives,
partitioned among $M$
distinct lengths, $l_m$, $m=1\cdots M$, upto some maximum value $L$,
with degeneracies $n_m$, $\sum_{m=1}^M n_m=N$. For sufficiently large $M$,
$L$, the sequence of lengths converges to a continuum and we can replace
the factor in square brackets in Eq.\ (\ref{eq:selb}) with an integral over
the density of lengths \cite{huber,mckean}:
\begin{eqnarray}
&&\lim_{L\to\infty} [ \sum_{\{ \delta ; l(\delta)<L \}} \cdots ~ ] =
   {{1}\over {\Gamma(s)}} \sum_{p=1}^{\infty}
   \nonumber \\
  &&\times \int_{l_{\rm min.}}^{L} dl n(l)
   2l e^{-pl/2}(1-e^{-pl})^{-1} I(pl;\epsilon,m^2).
\label{eq:asymp}
\end{eqnarray}
The growth in the degeneracy of primitive elements of fixed length $L$
is known from Huber's estimate \cite{huber,mckean}:
\begin{equation}
\lim_{L\to\infty} \int_{l_{\rm min.}}^{L} n(l) dl = (1 + \alpha) {{1}\over{L}}
   e^L ,
\label{huber}
\end{equation}
where $\alpha$ is a constant of order less than one. Interchanging the
order of integration over $t$, $l$ and substituting from Eq.\ (\ref{huber}),
we see that the $p=1$ term in the sum dominates the bound. Thus,
\begin{equation}
\lim_{L\to\infty} \zeta_R(s) \le {{1}\over {\Gamma(s)}}
L e^{-L/2} \cdot {{1}\over{L}} e^L \cdot I(L;\epsilon,m^2) .
\label{bzeta}
\end{equation}
The integral over $t$ can be bounded for large $L$ by splitting the
range of integration, $\int_\epsilon^{\infty} = \int_0^\infty -
\int_0^{\epsilon}$, and estimating the correction to the resulting modified
Bessel function from the short time behavior of $I(L;\epsilon,m^2)$. The
result is the bound:
\begin{equation}
\lim_{L\to \infty} \zeta_R(s) \le {{e^{L/2}}\over {\Gamma(s)}}[
     L^{\nu -1/2}e^{-L/2} + O( \epsilon^{\nu+1} L^{-2} e^{-L^2/4\epsilon})] ,
\label{bound}
\end{equation}
where $\nu=s-1/2$. The leading behavior can already be extracted from
\cite{dhokerphong1}. We have simply clarified its extension to the Greens
function on surfaces with boundaries. As expected, for large $L$,
the $\epsilon$ dependent correction
is insignificant unless $r$$<<$${\sqrt{\alpha'}}$. Similar estimates
can be made for the Greens function at the source, $G({\rm Cosh}(l_{\rm
min.};\epsilon))$, and the Greens functions of the Laplace-Beltrami
operators for higher rank tensor fields on $S_b$
\cite{fay,dhokerphong1,wolpert3}. These include the vector Laplacian
that appears in the measure for the string path integral. The functional
methods described here can be extended to amplitudes with general
boundary conditions \cite{polchinskivol1}: moving \cite{velocity},
mixed \cite{indextheo}-- for configurations with Dbranes of different
dimensionality, external field and finite temperature \cite{finitetemp},
and, most importantly, supersymmetric. Work in this direction is in
progress.

Thus far, we have restricted ourselves to a discussion of the open
string amplitude with multiple boundaries on {\em separated}
Dbranes, $r_{ij} $$>$${\sqrt {\alpha'}}$, for every $i,j$$=$$1$,
$\cdots$, $b$. Consider the configuration with $N$ superimposed
Dbranes, within distances of $O({\sqrt{\alpha'}})$, and a single
Dbrane at a distance $r$$>>$${\sqrt{\alpha'}}$. The
$l$$\to$$\infty$ asymptotics of the string path integral yields
the behavior of a nonabelian, $U(N)$, gauge theory on the Dbrane
stack \cite{polchinskivol1}.

\vskip 0.1in We conclude with the observation that modifying the
Weyl-covariant open string perturbation theory in stretched string
backgrounds by introduction of a finite world-sheet ultraviolet
cutoff, $\epsilon$ $\simeq$ $\alpha^{\prime 1/2}/r$, where $r$ is
the length of the stretched string, can indeed give a prescription
that preserves both open-closed world-sheet duality, and the
target space T-dualities. More importantly, it might also provide
a quantitative description of the crossover regime:
\begin{equation}
N^{1/7} M_P^{-1} ~<~ r ~<~ N^{1/3} M_P^{-1} ,
\label{eq:regime}
\end{equation}
between supergravity and perturbative super Yang-Mills theory
\cite{imsy,mlcone,itzhaki}. Here, $M_P^{-1}$$=$$g_c^{1/3}
{\sqrt{\alpha'}}$, is the eleven dimensional Planck length, and
$r$ has been identified with the typical length scale probed by
the dynamics, i.e., the spatial separation of two point sources in
11-d supergravity. This corresponds to D0brane dynamics in the
type IIA string theory or, by a T-duality in the direction $X^9$,
to D1brane dynamics in the type IIB string theory. The upper bound
in Eq.\ (\ref{eq:regime}) is the t'Hooft radius, marking the
transition to perturbative SYM theory. The lower bound, where the
closed string coupling $g_c$$\sim$$O(1)$, is the transition to the
low energy limit of M theory. But it should be emphasized that, as
a consequence of our having explictly broken worldsheet Weyl
invariance, this prescription is only {\em effective}: the
renormalizability of the low energy gauge theory limit in an
exactly solvable conformal field theory background of an infrared
finite perturbative string theory, even in the absence of target
spacetime supersymmetry \cite{holo}, has been obscured
\cite{zeta}.

\vskip 0.1in What is the significance of a world-sheet ultraviolet
regulator that preserves both open-closed world-sheet duality, and
T-duality, although it violates the Weyl invariance of the
covariant worldsheet formalism? Our conjecture is that IIA string
perturbation theory with finite ultraviolet cutoff
$\epsilon$$\sim$$O(\alpha'/r^2)$ might fill the gap between the
kinematic regimes of DLCQ, the discrete light cone quantization of
M theory, and of 11-d supergravity \cite{mlcone,itzhaki}. Upon
T-dualizing, this would cover the crossover region between the
supergravity and perturbative SYM regimes of the IIB string theory
\cite{malda,imsy}. Whether this can lead to detailed comparisons
between the modified open string perturbation theory, and
nonperturbative formulations of DLCQ \cite{itzhaki}, remains to be
investigated in the future.

\vskip 0.2in
\noindent{\large\bf Acknowledgments}
\vskip 0.1in

I would like to thank J. Polchinski for inspiration. I also thank
the Institute for Theoretical Physics and the Aspen Center for
Physics for their hospitality. I am grateful to E. D'Hoker for
providing me with references and comments on previous work.
Finally, I must thank my students E. Novak and Y. Chen for their
questions during the course of this work. This research is
supported in part by NSF grant PHY 97-22394.

\vskip 0.2in\noindent{\bf Note added (Jan 2005)}: Minor
corrections in wording have been made in the text of the paper to
clarify that the introduction of an explicit cutoff on the
boundary loop lengths, as might seem natural in the background of
stretched strings, amounts to a formal violation of worldsheet
Weyl invariance \cite{zeta}. We emphasize that the conclusion that
a perturbatively infrared finite string theory is automatically
ultraviolet finite, even in the absence of target spacetime
supersymmetry \cite{holo,micro}, holds only in the limit of
unbroken Weyl invariance: $\epsilon$ $\to$ $0$ \cite{zeta}. A
consequence of broken Weyl invariance is that the worldvolume
gauge theory obtained in the low energy limit of the string theory
is only a Wilsonian {\em effective} field theory, with ultraviolet
(uv) cutoff provided by the tension of the stretched string. From
the perspective of the gauge theory, the uv cutoff can be
interpreted as the mass of the Goldstone boson associated with the
spontaneous breaking of Poincare invariance in the bulk target
spacetime by the Dbranes. Refs.\ \cite{zeta,holo,micro} are new.

\appendix
\section{\bf The Selberg Trace Formula}

\vskip 0.1in
The basic observation underlying the derivation of Selberg's trace formula
is as follows \cite{selberg}. A readable introduction appears in \cite{mckean}.
The Greens function of the scalar Laplacian on $S$, with covering space the upper half
plane, $H$, is an example of a \lq\lq point--pair invariant ", i.e., a symmetric kernel
function $k(x,y)$, with $x,y$ any two points in some domain $\cal F$ of arbitrary
dimension, such that
$k(\gamma [x], \gamma [y])=k(x,y)$ for all $\gamma \in \Gamma$, the covering group
of $F$. Then the action of the group induces a symmetric kernel function
$\sum_{\gamma \in \Gamma} k(x, \gamma [y])$ in the covering space, $\cal C$, where
${\cal F}={\cal C}/\Gamma$. Selberg's trace formula expresses the kernels of
a family of invariant operators, $\{ {\cal L} \}$, on the domain $\cal F$ as
functions of group invariant objects, $\{\sigma(x,y)\}$. In our example,
these are the Laplace-Beltrami operators acting on arbitrary rank tensors
on $S$, and there is a single group invariant, $\rho_H(z,z')$, the hyperbolic distance
between the two points. The measure for the kernel functions, $\{k(\sigma(x,y))\}$,
is the group invariant measure on ${\cal C}$, in our example, the $PSL(2,R)$
invariant area element on $H$.

Selberg's trace formula expresses any kernel function on $\cal F$ as a group-theoretic
trace over a distinguished subset of elements of the covering group $\Gamma$ known as the
{\em primitive elements}: elements which cannot be expressed as a power of any other
element of the group \cite{selberg}. A crucial role is played in this derivation by
the similarity relation of the covering group. To illustrate this, it is helpful to
understand a concrete example and we will therefore begin with a review of the domain and
primitive elements of the modular group of $S$. But the reader should keep in mind
that the applicability of Selberg's method to the computation of symmetric kernel
functions is much more general than this specific example illustrates. For an account of
some physics applications other than string theory, consult the articles and references
cited in the collection \cite{appns}.

\section{Uniformization and Kleinian Groups}

\vskip 0.1in
Any Riemann surface $S_{(b,c)}^{h}$, with
$b$ holes, $c$ crosscaps, and $h$ handles, can be globally uniformized,
i.e., represented parametrically by single valued holomorphic or meromorphic
functions in a subdomain, $F_{(b,c)}^h$, of the Riemann sphere,
${\hat C}$, the universal covering space of $S$ \cite{bers}.
An isometry of the Riemann sphere is an element of the Mobius group, linear
fractional transformations on the complex plane, acting on coordinates, $z$,
as $\gamma [z] = \frac {a z+ b}{ cz+d}$, with $a,b,c,d$$\in $$\bf C$, and
$ad-bc=1$. The discrete subgroup of Mobius transformations associated with
the isometries of the subdomain $ F$, is a Kleinian
subgroup of the Mobius group. A special case is a Fuchsian subgroup,
obtained under uniformization to a subdomain of the Poincare upper half plane,
$H$. Every compact oriented surface with negative Euler character, $\chi$$=$$2-2h-b$,
constant Gaussian curvature, $R_g$, and geodesic boundaries, can be globally
uniformized to a polygon with $-9\chi$ sides in $H$, $F$$\sim$$H/\Gamma$,
whose shape is parameterized by $-3\chi$ real moduli parameters. The Gaussian
curvature is normalized to $R_g$$=$$-1$ and, from the Gauss--Bonnet theorem,
the area equals a constant, $A=-4\pi \chi$.

\vskip 0.1in
Geodesics in the complex plane with hyperbolic metric are circular arcs
degenerating, for sufficiently large radius, to straight lines perpendicular
to the real axis. Under a conformal mapping of $H$ to the interior of
the unit disk, $D$, the domain $H/\Gamma$ maps to a subdomain
of $D$, with homeomorphically related hyperbolic metric. This is an
equivalent global uniformization of the Riemann surface. Note that the Kleinian
covering groups associated with either choice of global uniformization
necessarily coincide, although the subdomains of the complex plane are distinct.
One can move freely between these two equivalent choices of domain as
convenient \cite{eigtext}.

\vskip 0.1in
It is possible to give an explicit representation in terms of PSL(2,R) matrices
for the generating elements of the Fuchsian group, $\Gamma_b$, of an orientable Riemann
surface, $S_b$, with $b$ holes and no handles \cite{keen}. The conjugacy classes
of $\Gamma_b$ correspond to homotopically inequivalent classes of closed paths on $S_b$.
Cutting open $S_b$ along $b-3$ interior geodesic paths gives $b-2$ {\em pants}:
surfaces of constant curvature $-1$ with three holes, $\cal P$.
Then the generating elements of $\Gamma_b$ can be expressed as
simple products of the generating elements of the individual $\Gamma^{(j)}_3$,
$j$$=$$1$, $\cdots$, $b-2$, as reviewed in \cite{abikov}\cite{keen}\cite{wolpert1}.

A pants surface has three homotopically inequivalent classes of paths, $C[i]$.
An arbitrary non-intersecting path on $\cal P$ is homotopic to one of 3 boundaries,
but $C[1]C[2]$$\sim$$C[3]$, so that the matrix
representatives of the generating elements satisfy the constraint, $\prod_{i=1}^3$$
\gamma_i$$=$${\bf 1}$. Consider an arbitrary path on $\cal P$ joining any two points,
$P$, $P'$.  Under uniformization to a subdomain of $H$, this maps into a semi-circular
arc of length parameter, $|{\rm tr} \gamma|$, where $z'$$=$$\gamma[z]$,
and $\gamma$ is an element of the covering group
$\Gamma_3$. The length parameter coincides with the geodesic length between the
points $z$, $z'$, on $H/\Gamma_3$ for those linear transformations that act as
{\em magnifications}. We will be especially interested in geodesic paths. These are
singled out
by an application of the similarity relation for the covering group: every element
of a Fuchsian group $\Gamma_b$ can be related by a similarity transformation to an
orientation preserving magnification, $z'$$=$$\alpha z$ \cite{abikov}. The
magnifications, $\{\alpha\}$, serve as representative elements of the conjugacy
classes of the covering groups, $\Gamma_b$.

\section{\bf Fundamental Polygons and Primitive Elements}
\label{prim}

\vskip 0.1in
Let us understand the meaning of a conjugacy relation for the elements of
the fundamental group $\Gamma$ of some domain ${\cal F}$$\sim$${\cal C}/\Gamma$.
As explained above, an arbitrary element in $\Gamma$ can be transformed
to a representative element, $\alpha$, by a conjugacy relation,
$\beta^{-1} \alpha \beta$$=$$\gamma $, for some $\beta \in \Gamma$.
The minimum distance along the curve $(z,\alpha[z])$, measured with the
group invariant metric, for paths belonging to the conjugacy class of $\alpha$
determines a length parameter. For the hyperbolic metric and the modular
group $\Gamma_b$, this parameter is:
\begin{eqnarray}
l(\alpha) &\equiv& {\rm Min} ~  {\rm Cosh}^{-1} \left [ 1 + \half
       {{|z-\alpha [z]|^2}\over{\alpha ({\rm Im} z)^2}} \right ]  \cr
          &=& {\rm Cosh}^{-1} [ \half (\alpha + \alpha^{-1}) ] \quad ,
\label{minl}
\end{eqnarray}
coincident with the trace of the representative group element,
$\alpha$$\in$$\Gamma$. Thus, the conjugacy class of the element $\alpha$
is isomorphic to a deformation class of geodesic paths: paths homotopically
equivalent to $\alpha [z_0]$, defined with respect to some base point $z_0$
that varies freely in $H/\Gamma_b$. The length parameter for repeated
magnifications is additive, giving the relation, $l(\alpha^n)=|n|l(\alpha)$,
for every $n \in Z$ \cite{mckean}. These comments can be generalized to the
general case.

\vskip 0.1in
This suggests a natural prescription for performing the sum over conjugacy
classes of a covering group $\Gamma$. We begin by defining a {\em primitive} element,
$ \delta $: a group element which cannot be expressed as a power of any other
element in $\Gamma$. From the relation above, an arbitrary magnification,
$\alpha$$\in$$\Gamma$, can be expressed as a power, $n$, of a primitive
magnification, $\alpha$$=$$\delta^n$. Arbitrary group elements, $\gamma$,
are then obtained by a sum over elements in the conjugacy class of
$\alpha$: $\gamma$$=$$\beta^{-1}
\delta^n \beta$, $\forall$ $\beta$$\in$$\Gamma$. This would overcount by an
overall infinity due to elements in $\Gamma$ that commute with $\delta$, since
for any such element the conjugacy relation maps $\delta^n$ into itself. Let
us therefore define the centralizer of the primitive element $\delta$,
$\Gamma^{(\delta)}$, as the set of elements in $\Gamma$
that commute with $\delta$. Running once through the primitive
magnifications, $\delta$, and then summing, for fixed $\delta$, over every
element $\beta \in \Gamma/\Gamma^{(\delta)}$, runs exactly {\em once} through
the conjugacy classes of the group, $\Gamma$. The conjugacy class of
any primitive magnification is therefore the set:
\begin{equation}
\{ \delta \} \equiv \{ \gamma ~|~ \gamma= ( \beta^{-1} \delta^n \beta)
~ \forall ~ \beta \in \Gamma/\Gamma^{(\delta)} , ~ n \in {\rm Z}^+ \} \quad .
\end{equation}
Thus, a sum over the conjugacy classes of the covering group can be written as
the nested sum:
\begin{equation}
\sum_{\{\delta\}} = \sum_{\delta} \sum_{\beta \in \Gamma/\Gamma^{(\delta)} }
                      \sum_{n=1}^{\infty}
    \quad .
\label{nest}
\end{equation}
This identity can be used to relate the Greens function on the covering space
$H$ to that on the fundamental polygon, $H/\Gamma_b$. This analysis of
primitive elements for a Fuchsian group straightforwardly
generalizes to the loxodromic Kleinian groups associated to a nonorientable
surface \cite{bers}. It should be noted that Selberg's formalism can be
applied to the Schottky
covering group of $S$, under global uniformization to a disjoint subdomain
of the flat complex plane: the fundamental domain of a genus $h$ closed
Riemann surface is the region in the complex plane exterior to $2h$
isometric circles, $C_j$,$C'_j$, $j=1, \cdots, h$. However, the
decomposition into primitive elements could get involved.

\section{\bf Derivation of the Trace Formula}

The analysis of conjugacy classes of the Fuchsian group
$\Gamma_{b}$ applies to any covering
group for which we can identify primitive elements.
For such a group,
the sum over conjugacy classes, $\sum_{\{\delta\}}$, can be written as
the nested sum:
\begin{equation}
\sum_{\{ \delta \}} \int_{{\cal C}/\Gamma} dx ~ k(x,\gamma [x]) =
     \sum_{\delta} \sum_{\beta \in \Gamma/\Gamma^{(\delta)} }
         \sum_{n=1}^{\infty} \int_{{\cal C}/\Gamma}  dx ~
              k(x, (\beta^{-1} \delta^n \beta) [x] )
\end{equation}
where $dx$ denotes the group invariant measure in the domain
$\cal F$$\sim$${\cal C}/\Gamma$.
Extracting the identity element from the sum over primitive elements,
we can simplify:
\begin{equation}
\sum_{\{\delta\}} \int_{{\cal C}/\Gamma} dx ~ k(x,\gamma [x]) =
     \sum_{\delta \neq {\bf 1}} \sum_{n=1}^{\infty}
      \sum_{\beta \in \Gamma/\Gamma^{(\delta)} }
        \int_{\beta[{\cal C}/\Gamma]}  dx ~ k_0(x, \delta^n [x] )
             ~+~ \int_{{\cal C}/\Gamma}  dx ~ k_0(x, x ),
\label{extractid}
\end{equation}
where $\beta[{\cal C}/\Gamma]$ denotes the image of the fundamental
domain under the action of the group element, $\beta$. The result is
therefore expressed as a sum over kernels $k_0(x,\delta^n [x])$, each
of which is integrated over the domain,
${\cal C}/\Gamma^{(\delta)}$, of an inconjugate primitive element $\delta$:
\begin{equation}
     \sum_{\delta \neq {\bf 1}} \sum_{n=1}^{\infty}
        \int_{{\cal C}/\Gamma_{\delta}}  dx ~ k_0(x, \delta^n [x] )
             ~+~ \int_{{\cal C}/\Gamma}  dx ~ k_0(x, x ) \quad .
\label{fundakernel}
\end{equation}
Here, $k_0(x,\gamma [x])$ is the kernel function induced on $\cal C$ by the
action of the covering group. Note that this mapping is bijective so we can
interchange the roles of $k(x,y)$ and $k_0(x,y)$, obtaining either one
from the other, given an explicit parameterization of $\Gamma$. The second
term in this equation is simply the area of the
domain multiplied by the kernel function with coincident
arguments: $Area({\cal C}/\Gamma) ~ k_0(\sigma=0)$. This is the Selberg trace
formula \cite{selberg}.

\section{Length Spectrum and Poincare Theta Series on $S_b$}

For the Fuchsian groups, $\Gamma_b$, the Selberg formula can be
simplified as follows. Consider a magnification, $\delta^p[z]=e^{pl(\delta)}z$.
Then,
\begin{equation}
{\rm tr}'e^{-(\Delta +m^2)t} =  \sum_{\delta \neq {\bf 1}} \sum_{p=1}^{\infty}
  \int_{H/\Gamma^{(\delta)}} {{d x d y}\over{y^2}}
                     {\hat G} (z, e^{p l(\delta[ z])};t )
 + A \cdot {\hat G}(z,z;t) \quad ,
\label{itee}
\end{equation}
where the argument of ${\hat G}(z,z';t)$ is the hyperbolic distance
between $z$, $z'$ in $H/\Gamma_b$:
\begin{equation}
{\rm Cosh}^2 (\rho_H (z,e^{pl(\delta)/2)}z)) = 1 + (1+x^2/y^2)
       {\rm Sinh}^2 (\half p l(\delta)) \quad ,
\label{magni}
\end{equation}
where the contribution from the identity element to the group theoretic trace
has been singled out. Here, $A$$=$$-4\pi\chi$. A change of variables,
$(x,y)$$\to$$(u$$=$$x/y,l'$$=$${\rm ln}y)$, in the trace formula gives
the result:
\begin{equation}
{\rm tr}' e^{-(\Delta+m^2)t} =  \sum_{\delta \neq {\bf 1}}
         \sum_{p=1}^{\infty} \int_{-\infty}^{\infty}  du
           \left [ {{ {\sqrt{2}} l(\delta) }\over{(4\pi t)^{3/2}}}
   \int_{\rho_H}^\infty db {{ b e^{-b^2/4t - M^2 t} }\over{
     {\sqrt{ {\rm Cosh} b - {\rm Cosh} \rho_H }} }} \right ] ,
\label{uncouple}
\end{equation}
where we have suppressed the
contribution to the trace from the identity element:
$A \cdot {\hat G}({\rm Cosh}(l_{\rm min.});t)$, the area times the Greens
function at the source.
In the absence of an ultraviolet cutoff on the world-sheet, we can
simply replace $l_{\rm min.}$ with zero but it should be kept in mind
that $l_{\rm min.}$ is really of $O(\alpha'/r^2)$, world-sheet ultraviolet
regulator, $\epsilon$, dependent.

Interchanging the orders of integration over $b$,$u$, the half-space,
$b\ge \rho_H$, with $u$ arbitrary, can be mapped to the half-strip,
$-u_0\le u \le u_0$, $b\ge pl(\delta)$, where
$u_0$$=$$ [{\rm Cosh}(b)-{\rm Cosh}(pl(\delta))]^{1/2}/{\sqrt{2}}
{\rm Sinh}(pl(\delta))/2$. Performing the integrations
gives the Poincare theta series:
\begin{equation}
\Theta(t)= \left [ \sum_{\delta \neq {\bf 1}} \sum_{p=1}^{\infty}
             {{l(\delta)/2}\over{{\rm Sinh} (pl(\delta)/2) }}
               e^{- p^2 l^2/4t} \right ]
                {{e^{-M^2t }}\over{(\pi t)^{1/2} }}
\quad .
\label{resultit}
\end{equation}

\section{Selberg Zeta Function}

Selberg's trace formula can be converted to infinite product form,
yielding the Selberg zeta functions with properties similar to those
of the Jacobi theta functions. These are especially helpful in
examining the large and small $l(\delta)$ asymptotics. A Laplace
transform of the Poincare theta series, $\Theta(t)$, is an entire
function in the $s$ plane and, in particular, is regular at the
origin, $s=0$ \cite{mckean}. Consider the logarithmic derivative
of $Z(s)$ with respect to the variable $s$. Then,
\begin{equation}
{{Z'(s)}\over{Z(s)}} \equiv (s- \half ) \int_0^{\infty} e^{-s(s-1)t}
           \Theta (t) dt \quad ,
\label{zetas}
\end{equation}
which can be considered as a functional equation whose solution is
the Selberg zeta function, $Z(s)$, defined as:
\begin{equation}
Z(s) = \prod_{\delta \neq {\bf 1}} \prod_{k=0}^{\infty} [
1- e^{-(s+k)l(\delta)} ] \quad .
\label{prod}
\end{equation}
It is easy to verify that this definition is consistent with the
integral transform of
$\Theta(t)$. We begin with the left-hand-side:
\begin{eqnarray}
{{Z'(s)}\over{Z(s)}} &=& \sum_{\delta \neq {\bf 1}} \sum_{k=0}^{\infty}
           l(\delta) {{e^{-(s+k)l(\delta)}}\over{1-e^{-(s+k)l(\delta)}}}
               \cr
       &=&  \sum_{\delta \neq {\bf 1}} l(\delta) \sum_{k=0}^{\infty}
                      \sum_{n=1}^{\infty} e^{-n(s+k)l(\delta)} \cr
       &=&  \sum_{p=1}^{\infty} \sum_{\delta \neq {\bf 1}}
                  {{l(\delta)/2}\over{{\rm Sinh}(\half l(\delta^p))}}
                         e^{-(s-\half)l(\delta^p)} \quad ,
\end{eqnarray}
where we have used $l(\delta^p)=pl(\delta)$. We recognize the argument
of the sum as $\Theta(t)$ by making use of the following
identity for Gaussian integrals:
\begin{equation}
e^{-(s-\half)l(\delta^p)} ~=~
        (s-\half) \int_0^{\infty} {{dt}\over{(4 \pi t)^{1/2}}}
               e^{-(s-\half)^2t - |l(\delta^p)|^2/4t}
                 \quad .
\label{gaussian}
\end{equation}
Direct substitution yields the functional relation above:
\begin{equation}
{{Z'(s)}\over{Z(s)}} = (s-\half) \sum_{p=1}^{\infty} \sum_{\delta \neq {\bf 1}}
     {{2 l(\delta)}\over{{\rm Sinh}(\half l(\delta^p))}}
         \int_0^{\infty} {{dt}\over{(4 \pi t)^{1/2}}}
               e^{ - |pl(\delta)|^2/4t} e^{-t/4 -s(s-1)t}
                 \quad .
\label{verify}
\end{equation}

\end{document}